%% file: ltexpprt.tex
\documentclass[twoside,leqno,twocolumn]{article}
\usepackage{ltexpprt}

\usepackage{bbm}
\usepackage{multirow}
\usepackage{graphicx}
\usepackage{subcaption}
\graphicspath{{./Images/}}
\usepackage{amssymb}
\usepackage{enumitem}
\usepackage{color}
\usepackage{url}
\usepackage{amsmath}
\usepackage{algorithm}
\newtheorem{definition}{Definition}

\begin{document}

\title{\Large First Study on Data Readiness Level}

\author{
Hui Guan\thanks{Department of Electrical and Computer Engineering, North Carolina State University. Email: \{hguan2, ahk\}@ncsu.edu} 
\and
Thanos Gentimis\thanks{Department of Mathematics, Florida Polytechnic University. Email: agentimis@flpoly.org}
\and
Hamid Krim\footnotemark[1] 
\and
James Keiser\thanks{ Laboratory for Analytic Sciences. Email: keiserjm@gmail.com }
}
%\title{\Large First Study on Data Readiness Level}
\date{}

\maketitle

%\pagenumbering{arabic}
%\setcounter{page}{1}%Leave this line commented out.

%\begin{abstract} \small\baselineskip=9pt This is the text of my abstract. It is a brief
%description of my
%paper, outlining the purposes and goals I am trying to address.\end{abstract}

\begin{abstract} \small\baselineskip=9pt We introduce the idea of Data Readiness Level (DRL) to measure the relative richness of data to answer specific questions often encountered by data scientists.
        We first approach the problem in its full generality explaining its desired mathematical properties and applications and then we propose and study two DRL metrics. Specifically, we define DRL as a function of at least four properties of data: Noisiness, Believability, Relevance, and Coherence. The information-theoretic based metrics, Cosine Similarity and Document Disparity, are proposed as indicators of Relevance and Coherence for a piece of data. The proposed metrics are validated through a text-based experiment using Twitter data.
\end{abstract}

%\section{Problem Specification.}In this paper, we consider the solution of the $N \times
%N$ linear
%system
%\begin{equation} \label{e1.1}
%A x = b
%\end{equation}
%where $A$ is large, sparse, symmetric, and positive definite.  We consider
%the direct solution of (\ref{e1.1}) by means of general sparse Gaussian
%elimination.  In such a procedure, we find a permutation matrix $P$, and
%compute the decomposition
%\[
%P A P^{t} = L D L^{t}
%\]
%where $L$ is unit lower triangular and $D$ is diagonal.

\input{Introduction}

\input{DRL}

\input{Background}
\input{Definitions}
\input{Methodology}

\input{Experiments}
\input{Relatedwork}
\input{Conclusion}
\input{Acknowledgement}

\bibliographystyle{unsrt}
\bibliography{bib}

\end{document}

%% file: Introduction.tex
\section{Introduction}

\label{sec:introduction}

 % background and motivation

Data nowadays are produced at an unprecedented rate; Cheap sensors, existing and synthesised datasets, the emerging internet of things, and especially social media, make the collection of vast and complicated datasets a relatively easy task. The era of big data is obviously here but unfortunately without an equal advance in the science of understanding it yet. With limited time and human power, the ability to effectively harness the power of big data is a problem encountered by many companies and organizations. Enormous amounts of data take up too much storage and computing resources. That, however, does not necessarily mean an increase of valuable information or better actionable items. Only small amounts of data may address questions due to noise, redundancy, and non-relevance. 

To increase effectiveness in data storage and handling, companies and organizations are focusing on driving robust data analysis techniques, such as Robust Principal Component Analysis \cite{candes2011robust} and K-medoids \cite{kaufman1987clustering}\cite{frey2007clustering}, to extract insights from data and downsize storage by keeping only the relevant information. However, the general theme of ``garbage in, garbage out" still applies, and with big data, the problem becomes even more pronounced. For example, no meaningful patterns would be recognized if the data itself doesn't contain much information, or even worse, phantom patterns will appear if the data is not relevant.   

% goal
Metrics that can evaluate the sufficiency, effectiveness and value of the collected data will bring incalculable benefits on selecting valuable data sets to analyze. Not only will the unnecessary cost of collecting redundant data be reduced, but also the efficiency of obtaining insightful results will be improved. The Data Readiness Level (DRL) measurement was first proposed by the Laboratory for Analytic Sciences at NC Sate University as a means to quantify that relevance. As the name suggests, the goal of DRL is to measure the relative readiness/richness of data to answer specific questions by various techniques.
%, essentially, a Signal-to-Noise Ratio (SNR) for data science. 

% Method
DRL should be a generic measure applied to a variety of data modalities, with inference/decision/answer to a question as a shared goal. Specifically, DRL should be a function of at least four properties of data: Noisiness, Believability, Relevance, and Coherence. In this paper, we focus on data sets comprised  primarily of documents since this is one of the most common datasets available and the literature in the field is rich enough to exploit various techniques. We assume that a set of documents is given to help answer a specific question. That dataset may contain relevant and irrelevant information and it does not have to be structured. The goal is to help guide a data scientist with a limited access to the entirety of the data corpus or alternatively with a limited time to analyze it, to quantitatively assess the capacity of a subset of that corpus to answer the question. This will be achieved by developing a computational measure to reflect the readiness of available data to answer such a question. 

This is, to the best of our knowledge, the first time this notion of ``goodness of unstructured data'' has been addressed in the context of data analytics for seeking answers to specific questions. We discuss here an unsupervised approach to computing DRL metrics on a wide variety of data. Moreover, we demonstrate its successful application to a collection of tweets. For computational efficiency and tractability, we use the assumption of ``Bag-of-Words'' (BOW)  in the case study on tweets. In the BOW model, a text is represented as the collection of its words, disregarding grammar and word order but keeping multiplicity \cite{harris1970distributional}. To that end,  topic modeling \cite{blei2012probabilistic}, often used in text mining and natural language processing as a dimension reduction, is the adopted strategy in our work. The Latent Dirichlet Allocation  approach \cite{Blei:03} was selected, but any other approach (e.g. %Term Frequency-Inverse Document Frequency (TF-IDF) \cite{manning2008scoring}, 
Latent Semantic Indexing \cite{hofmann1999probabilistic}, or Non-negative Matrix Factorization \cite{lee2001algorithms}) could have   been used, just as well. Cosine Similarity and Document Disparity are computed to measure Relevance and Coherence properties for a specific text collection. The underlying assumption is that a relevant set of documents should have high similarity to the question and low internal disparity, indicating high Relevance and high Coherence respectively. Our formulation is fully data-driven, hence more ``natural'' and better suited to the underlying structure of the data, and the proposed  metrics are relatively easy to compute.

%In this paper, our contributions are the studies of DRL metrics from an information-theoretic perspective.  We defined {\it cosine similarity} as an indicator of Relevance  and {\it document disparity} as an indicator of Coherence for a piece of data. 

% structure of the rest
The rest of this paper is organized as follows. In section \ref{DRL}, we discuss various required properties for a sound  DRL formulation and several relevant factors in its definition. In section \ref{Prereq} we provide an overview of the  mathematical, statistical and machine learning relevant tools we used. In sections \ref{definitions} and \ref{methodology}, we describe the theoretical framework of our proposed method and the methodology for implementing the framework.  For validation as well as illustrative purposes, we present a practical example on tweets in section \ref{experiment}. Some of the related work is described in the section \ref{Relatedwork} followed by a conclusion section \ref{conclusion}.

%% file: DRL.tex
\section{Definition of DRL}\label{DRL}

DRL is a function on a data set, which has been subjected to a sequence  of transformations  towards answering a query. In that sense, it reflects a {\em degree of maturity} towards accomplishing a target task.
As a metric, it may be evaluated at various stages of transformation of data, and  will  hence reflect the efficacy of each of  the various stages/analytics, or in one shot for the entire flow,  with a goal of refining and distilling the data more closely to successfully resolve the query.\par
As a valid and useful measure,  DRL has  to satisfy the following properties: 
\begin{enumerate}
\vspace*{-0.5\baselineskip}
\setlength\itemsep{-0.3em}
\item Easy computability. This ensures that DRL remains more advantageous than actually carrying out the information analysis. 
\item Stability and continuity. This will imply that any two close formulations (data, question, analytics) will yield similar DRL values. 
\item Scalability. This will safeguard its viability with increasing data size.
\item A maximally unsupervised property, but with the allowance of fine tuning the parameters.
\item A discriminative property, to make it useful for a data scientist in reaching a decision about the data at hand.
\vspace*{-0.5\baselineskip}
 \end{enumerate}
These properties are intrinsic to our proposed techniques and, up to minor adjustments,  can be shown to be stable under perturbations/changes of the input data, in our case, documents. The mathematical development of our DRL will  further unveil that the following have to be accounted for:
 
\begin{enumerate}
\vspace*{-0.5\baselineskip}
\setlength\itemsep{-0.3em}
\item \textbf{Data.} Given the application scope of DRL, diverse  data modalities may be addressed and any proposed metric or framework must address a variety of data. The data may  be noisy and unstructured.

\item \textbf{Analytics.} Given the great diversity of analytics/transformations and their specificity to problems, DRL may be operating along a number of possible dimensions, including time, cost, and computational complexity.  
%In addition to actual programs and automated processes, analytics should include individuals with specific skills, like translators, cryptographers, or in general, experts in a certain field. Operationalizing DRL will require synthesizing analytics and tracking down common paths and emerging trends and should lead to the ability to calculate bounds on the minimum and maximum ``improvement'' on the state of data one analytic can infer.
\item	\textbf{Objective Question (Question of Interest).} The nature of the proposed question may vary widely. Thus DRL must be flexible enough to accommodate binary, quantitative and/or probabilistic questions and answers. Additionally, questions may vary in substance and detail, and hence emanate from a population with a certain distribution, so an associated DRL will accordingly be in some sense a ``conditional'' quantity relative to the query.
 
\end{enumerate}

%In this report, we describe some avenues we have explored with two proposed DRL measures, and their derivation and evaluation relative to a data set with an objective question. Our experimental results will be described in section \ref{Con}.

To further clarify the problem statement, we focus on the application of DRL to unstructured text data. To that end, we assume that a corpus of documents is given to help answer a specific question. The goal of DRL is to quantitatively describe the amount of valuable information contained in the set of documents related to the question. We consider four dimensions of the set of documents in computing DRL: 
\begin{enumerate}
\vspace*{-0.5\baselineskip}
\setlength\itemsep{-0.3em}
	\item \textbf{Relevance:} the extent to which data is applicable and helpful for the question at hand. 
	\item \textbf{Coherence:} the extent to which data is consistent in subjects or topics.
	\item \textbf{Believability:}  the extent to which data is regarded as true and credible.
	\item \textbf{Noisiness:} the extent to which data is clean and correct.
	\vspace*{-0.5\baselineskip}
\end{enumerate}
The dimensions can be modified or discarded depending on the application. For example, for time-sensitive questions, the timeliness dimension should also be considered. In this paper, we account for only these four dimensions of data to make DRL as generic as possible. 

Mathematically, let $\mathcal {D}$ be the space of documents and let $D$ be a finite set of documents of interest with cardinality $N$, i.e.
$D=\{ d_1,\cdots, d_N\} \subset \mathcal{D}$, where $ d_i$ corresponds to the $i$-th document. Similarly, let $\mathcal{Q}$ be the space of questions and let $\mathbf{q} \in \mathcal{Q}$ be a specific question. Our goal is to compute $DRL(D|\mathbf{q})$.
The connection between the space of documents and that of questions is obvious if we treat questions as micro-documents. In this sense $\mathbf{q}\in\mathcal{Q} \subset \mathcal{D}$. 

Ideally, DRL should be a function that maps any set of documents to some fixed numerical range, for example, $[0,1]$, with 0 indicating no value at all and 1 indicating the most valuable data: $f: 2^{\mathcal{D}}\rightarrow [0, 1]$. An application of this would be to focus on the documents with the maximum possible DRL. Given the sets of documents $\{D_1, D_2, \cdots, D_M\}$, it is reasonable to only require the property: $DRL(D_i|\mathbf{q})>DRL(D_j|\mathbf{q})$ i.e. $D_i$ is more valuable than $D_j$ on answering the question $\mathbf{q}$. %An intuitive way to construct the function is to rank the set of documents based on the linear sum of the ranking on each dimension. 

%\begin{equation}
%DRL({D}|\mathbf{q}) = f({\mathcal N}, {\mathcal R}, {\mathcal B}, {\mathcal C}),
%\end{equation}
%where, ${\mathcal N}$, ${\mathcal R}$, ${\mathcal B}$ and ${\mathcal C}$ are noisiness, relevance, believability and coherence respectively. %This is a general framework on computing DRL on text.  I removed this part to keep the notation as clear and concise as possible. Also, we did not make any special analysis on the ideas you mention above.

%This measure should be depended on the four dimensions we described earlier. For example, Noisiness in text could mean the extent to which the document is well-written in grammar. Grammatically well written documents are good for current NLP tools. On the other hand tweets are much  ``noisier" than scientific papers because the former may contain unusual characters, symbols, grammar and specific words or even slang language. Tweet sentences may also not follow linguistic rules making it hard for automated text pre-processing tools like the Part-Of-Speech tagging and parsing. Noisiness in text could also mean the extent to which the document explains the relevant content. In this case a document  with a reasonable discourse structure is easier to read and thus ready to answer the relevant question.  For our paper, believability was dropped assuming that all documents are of the same level of believability but in general it should be factored in the measurements. 

%% file: Background.tex
\section{Background}\label{Prereq}

To cast our proposed DRL methodology in an applied setting, and keep this report self-contained for a document-based experimental evaluation,  we provide some brief background on text mining, making connections with the relevant research in the area.

\subsection{Latent Dirichlet Allocation}

Since our DRL experimental validation involves document analysis,  it is  natural to approach documents as bags of words, as often done in classic data mining. The size of a given corpus of documents may, however, be prohibitively large making this approach computationally expensive and negatively impacting the efficiency of the DRL computation. To alleviate that problem we have adopted some tools from topic modeling and specifically the concept of Latent Dirichlet Allocation \cite{Blei:03}. We assume thus that the question and the document that are addressing the same topics are likely to be relevant and use that as a bottom line for our relevance measure.

The essence of LDA is that documents are distributions over topics, the latter being themselves probability distributions over words. Suppose that ${D}=\{{\mathbf d}_1,\cdots, {\mathbf d}_{N}\}$ is a set of $N$ documents. Let $K$ be the number of topics and $V$ be the number of words in the vocabulary. Let $\boldsymbol{\varphi}_k\in \mathbf{R}^V, k = 1,\cdots, K$ be the distribution of words corresponding to topic $k$. Given a document ${\mathbf d}_i\in D$ , let $N_i$ be the number of words and $\boldsymbol{\theta}_i\in \mathbb{R}^K$ be the distribution of topics. The generative probabilistic model defined in LDA is as follows:
\begin{enumerate}
\vspace*{-0.5\baselineskip}
\setlength\itemsep{-0.3em}
	\item Choose $\boldsymbol{\theta}_i \sim Dir(\boldsymbol{\alpha})$ $i \in \{1, \cdots, N\}$, where  $Dir(\boldsymbol{\alpha})$ is the Dirichlet distribution with parameter $\boldsymbol{\alpha} = \{\alpha_1, \cdots, \alpha_K\}$.
	\item Choose $\boldsymbol{\varphi}_k \sim Dir(\beta)$ $k \in \{1, \cdots, K\}$ where, $Dir(\beta)$ is the Dirichlet distribution with parameter $\beta$.
	\item For each word $w_{i,j}$ in $j^{th}$ position of the $i^{th}$ document, where $i\in \{1, \cdots, N\}, j\in \{1, \cdots, N_i\}$:
	\begin{enumerate}
		\item Choose a topic $z_{i, j} \sim \text{Multinomial}(\boldsymbol{\theta}_i)$,
		\item Choose a word $w_{i, j} \sim \text{Multinomial}(\boldsymbol{\varphi}_{z_{i, j}})$.
	\end{enumerate}
\end{enumerate}

%We use LDA to identify the topics of the question and documents and check whether they are similar to each other with their topic representations. The more similar they are, the more relevant the dataset is to the question. 

%The essence of LDA is that documents are nothing but a mixture of topics, the latter   themselves  being   probability distributions over words. LDA uses the framework of a generative probabilistic model of data, with hidden variables representing the topic structure. The generative process may be described by a joint distribution of the hidden as well as the observed variables, which in LDA are the words in the documents.  Using Bayesian inference, the analysis is carried out with the help of this  joint distribution to result in a posterior  distribution.  The goal is of this posterior is to unveil  the hidden topic structure. 
%Common  assumptions made with  this model, include,
%\begin{enumerate}
%\item The total number of topics $K$ is assumed known and fixed.
%\item The word probabilities are parameterized by a $V\times K$ matrix $\boldsymbol{\beta}$, where $\boldsymbol{\beta}=(\boldsymbol{\beta}_1, \boldsymbol{\beta}_2, \dots, \boldsymbol{\beta}_K)$, which is what we need to estimate.
%\item The Dirichlet distribution is used as a prior for both the document-topic proportions $\boldsymbol{\theta}_d$ and the topic-word distributions $\boldsymbol{\beta}_k$.  Other distributions may, however, also be used as the prior. For example, in Correlated Topic Models, topics proportions are drawn from a logistic normal rather than a Dirichlet.
%\end{enumerate}

\subsection{The Semantic Space of Queries}

%Recall  a query, in document analysis, is a text based entity and the associated analytics of interest are common in text mining and natural language processing.
%Given, for instance, a set of news articles with a related question such as  ``What happened on $9/11$?", and our goal is to define a pertaining DRL. 

To better contextualize DRL conditioned on a question $\mathbf{q}$, a class of questions $Q$ conveying a similar ``message'' with possibly different words, is sought. This may be thought of as defining a semantic equivalence class to $\mathbf{q}$.  Let again  $\mathcal{D}$  be the set of all documents, each considered as a  bag of words using a high dimensional lexicon. %\cite{Kon:06}.  
Each question, regarded as a mini document, would correspond then to a representation in the \textit{Semantic Space} denoted by $\mathcal{S}$.
This space may roughly be associated to the space of distributions over topics, which, recall, are the result of the LDA turning documents (a bag of words) into distributions over topics. Thus what we should be basing our DRL on is the set of questions which have the same image under some function: $g:\mathcal{D}\to \mathcal{S}.$
The idea of this semantic space can be easily extended to other modalities, so long as a good proximity measure is identified. Unfortunately, such a measure is still fairly not well understood just like the notion of this space itself. 

To proceed with the mathematical development,  denote by $D$ a  family of documents which is a subset of $\mathcal{D}$. The LDA approach thus defines a function, 
\begin{equation}
\label{fun}
g_D:\mathcal{D}\to \mathcal{S},
\end{equation}
hence  associating with  each document its semantic meaning. A question $\mathbf{q}$ is typically  also projected onto this Semantic Space through the same function $g_D$. By construction this function $g_D$ depends on the document set itself, and hence the corresponding index $D$. It would be ideal if a general function could be found to attribute meaning to documents without the use of the specific setting; however, $g_D$ is good enough for our purpose. 

\begin{figure}[h!]
\vspace*{-0.5\baselineskip}
	\centering
	\includegraphics[width=0.8\columnwidth]{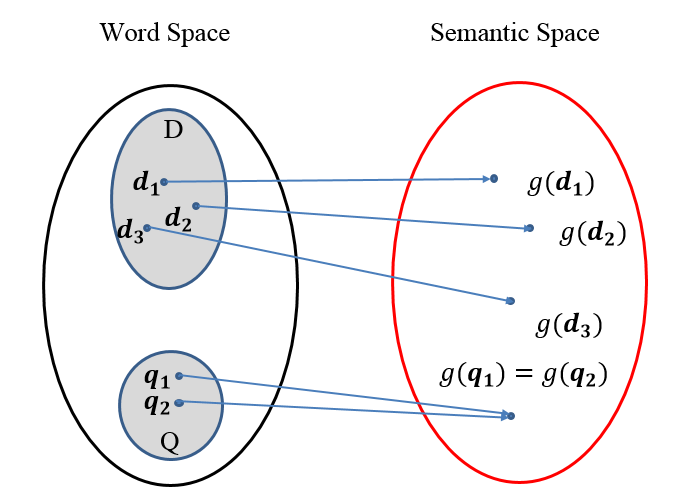}
	\vspace*{-0.5\baselineskip}
	\caption{Space mapping}
	\label{fig:space_mapping}
	\vspace*{-0.5\baselineskip}
\end{figure}

As shown in figure \ref{fig:space_mapping}, the documents $\mathbf{d_i} \in D \subset \mathcal{D}, i = 1,2,3$ are projected onto the semantic space, and  $\mathbf{q_1}, \mathbf{q_2}$ are two queries in $\mathcal{Q} \subset \mathcal{D}$, with the same representation in $\mathcal{S}$. The ``distance" between the query and the documents should be computed in terms of distances in the semantic space. 
In so doing, the DRL will depend on the ``meaning" of the question posed,  and avoid being indicative of the specific words used to formulate it. 

We note, that a philosophical analogy may be drawn between this formulation of the DRL with that of Information Retrieval(IR) as  related to  question expansion, which invokes techniques of  finding synonyms, stemming, spelling correction, etc.. We maintain that DRL should, however, go further and deeper by breaking down the concept for better understanding of the question, and for discovering the intent behind the question. 

%Presented with a set of words, an individual may associate a number of topics to the question at hand. Specifically,  given the question:``What happened on $9/11$?", one may very well associate  keywords such as ``terrorist'', ``attack'' and ``building'', which in turn may yield other relevant topics. An open question in this direction concerns a principled and algorithmic way of generating such related topics. Its computational complexity is the first emerging issue, and a conjecture of K-D trees solution is noteworthy at this point. In fact, this  difficulty lies at the center of a clear definition of  the Semantic Space, and  so the issue of  {\em semantic expansion} of a  question is an important future topic of  research.

\subsection{Jensen-Renyi Divergence}

Information theory and statistics are rich with measures of closeness (conversely divergence) between probability density functions (PDF). On the other hand, a PDF, itself, is a reflection of intrinsic behavior of a given population. It then makes eminent sense to seek the same notion of the concentration of data in a given population, across populations. That is precisely what is reflected in information theoretic measures such as divergence (for example, Kullback-Liebler (KL) divergence).
To that end, we build on some of our previous work in developing measures across an arbitrary number of PDFs , as a step beyond the well known and classical KL divergence between two PDFs. Viewing hence all documents as  distributions over topics using the LDA technique, we proceed to first define  the so-called Jensen-Renyi (JR) divergence \cite{He:03}.

\begin{definition}
	Let ${\bf p}_1, {\bf p}_2, \dots, {\bf p}_n$ be $n$ probability distributions on a finite set ${x_1, x_2,\dots, x_k}$, $k\in \mathbf{N}$. Each probability distribution is ${\bf p} =(p_1, p_2, \dots, p_k)$, $\sum_{j=1}^{k}p_j=1, p_j = P(x_j)\geq 0$. Let ${\boldsymbol \omega} = (\omega_1, \omega_2, \dots, \omega_n)$ be a weight vector and $\sum_{i=1}^{n}\omega_i=1, \omega_i\geq 0$. The JR divergence is defined as:
	\begin{equation}
	JR_\alpha^{\boldsymbol \omega}({\bf p}_1,\dots,{\bf p}_n) = R_\alpha \left( \sum_{i=1}^{n}\omega_i{\bf p}_i\right)-\sum_{i=1}^{n}w_iR_\alpha({\bf p}_i),
	\end{equation}
	where $R_\alpha({\bf p})$ is the Renyi entropy defined as:
	\[
	R_\alpha({\bf p}) = \frac{1}{1-\alpha}\log\sum_{j=1}^{k}p_j^\alpha, \alpha > 0 , \alpha \neq 1.
	\]
\end{definition}
The JR divergence is a convex function over ${\bf p}_1, {\bf p}_2, \dots, {\bf p}_n$ for  $\alpha \in (0,1)$. It achieves a minimum value of zero when  ${\bf p}_1, {\bf p}_2, \dots, {\bf p}_n$ are equal for all $\alpha > 0$ and gets its maximum value when ${\bf p}_i = {\delta_{ij}}$, where $\delta_{ij} = 1$ if $ i=j$ and $0$ otherwise. Many more properties of JR divergence can be found in \cite{He:03}.

\subsection{Sensitivity and Closeness}\label{sens}

In  numerical linear algebra and linear system theory, a non-singular matrix $A$ is ill-conditioned if a relative small change in $A$ can cause a large relative change in $A_{-1}$ : $(||x-x'||)/||x|| \leq k*{||B||/||A||}$, where $B$ is the perturbation input to the system $A$ and $k$ is the condition number $k = ||A||*||A^{-1}||$ \cite{belsley2005regression}. This concept is extended to include non-linear systems, like the LDA, using ideas from functional analysis on metric spaces as follows:

\begin{definition} Given a function $f:X\to Y$ where $(X,d_x)$ and $(Y,d_x)$ are metric spaces, we say that $f$ is $r$-locally $l$-Lipschitz if and only if for all $x\in X$ and for all $y\in X$ such that $d_X(x,y)\leq r$ we have that: \[d_Y(f(x),f(y))\leq l \cdot d_X(x,y).\]
\end{definition}

The idea of a Lipschitz constant $l$ is connected to the derivative of a function at a point (recall that a differential invokes a denominator as  $d_X(x,y)$). 
If $l$ is smaller than $1$ we say that $f$ is a contraction or in other words makes small errors even smaller (stabilizes). If the two spaces, however, have wildly different metrics, the Lipschitz constant is not very informative. To alleviate this problem we normalize the two metrics and define a new sensitivity number as follows:

\begin{definition} (\textbf{Sensitivity Number}).
	Given a function $f:X\to Y$ where $(X,\|\cdot\|_X)$ and $(Y,\|\cdot\|_Y)$ are spaces with norms, we define the relative $r$-sensitivity at the point $x_0$ to be: 
	\begin{equation}
	\label{eq:sensitivity_number_definition}
	\begin{split}
	s_1(x_0)= & \inf \{\frac{\|f(x)-f(x_0)\|_Y}{\|f(x_0)\|_Y}\div \frac{\|x-x_0\|_X}{\|x_0\|_X} |\\
	&  x\in X,\,\|x-x_0\|_X\leq r \} .
	\end{split}
	\end{equation}
\end{definition}

We will use these notions in Eq.\ref{eq:sensitivity_number_definition} to report a form of stability to small perturbations for our measures. 

%% file: Definitions.tex
\section{Theoretical Framework}
\label{definitions}

%According to Equation \ref{eq:DRL_definition2} in Section \ref{DRL}, we simplify DRL as a function of Noisiness ${\mathcal N}$, Relevance ${\mathcal R}$ and Coherence ${\mathcal C}$ dimensions of a given dataset. Here we give some definitions in our framework. 

In this section, we will define the two preliminary measures we propose for a functional DRL that correspond to the relevance of a set of documents to a certain question and its overall coherence.

\begin{definition} (\textbf{Relevance} ${\mathcal R}$).
	Let $M$ denote the number of sets of documents and $N_i$ denote the number of document in the $i$-th collection. Each document ${\bf d}_{j}^{(i)}$ is represented as: ${g(\textbf d}_{j}^{(i)})=[\theta_{j1}^{(i)},\theta_{j2}^{(i)},\dots,\theta_{jK}^{(i)}]$, $i=1,2,\dots,M, j=1,2,\dots,N_i$, where $\theta_{jk}^{(i)}$ is the proportion of topic $k$ in the $j$-th document on the $i$-th collection, using eq.(\ref{fun}). The question is also represented by ${g(\bf q)}=[\theta_1,\theta_2,\dots,\theta_K]$. Define then a Relevance metric, to be the cosine similarity measure between each document in the collection $D_i$ and the question $\bf q$:
\begin{equation}
	Sim(D_i, {\bf q}) = \frac{1}{N_i}\sum_{j=1}^{N_i}\frac{g({\bf d}_{j}^{(i)})\cdot g({\bf q})}{\|g({\bf d}_{j}^{(i)})\|\cdot\|g({\bf q)\|}}.
	\label{eq:relevance}
	\end{equation}
	\label{relevance}
\end{definition}
\vspace*{-0.5\baselineskip}

Formally, let  $\mathcal{X}$ be  a  collection of sets of documents and $\mathcal{Q}$ be a set of questions, the function, $Sim:\mathcal{X} \times \mathcal{Q} \to \mathbf{R}^+$ is a DRL measure which denotes the relation between each set and a question. Note that this may just as well  be applied to individual documents (when the document set size is 1) or  to the whole corpus of documents.
Eq.(\ref{eq:relevance}) obviously  induces a relation between two document sets as follows:
\begin{definition}
	Let $D_1$ and $D_2$ be two sets of documents and $\bf{q}$ an associated question as described above. We say that $D_1$ and $D_2$ are \textit{informationally equivalent relative to the question $\bf{q}$} if and only if \[ Sim(D_1,\textbf{q})=Sim(D_2,\textbf{q}).\]
\end{definition}
It is hard to find informationally equivalent document sets in practice, hence warranting a definition of  $\delta$-equivalent documents as follows:
\begin{definition}
	Let $D_1$ and $D_2$ be two sets of documents and $\bf{q}$ an associated question  as described above. We say that $D_1$ and $D_2$ are \textit{$\delta$-informationally equivalent relative to the question $\bf{q}$}, if and only if \[|Sim(D_1,\textbf{q})-Sim(D_2,\textbf{q})|\leq \delta. \]
\end{definition}

If we consider the set $\mathcal{X}$ under the informational equivalence relation, or the $\delta$ informational equivalence, then the function $Sim$ induces a total order relative to the question $\textbf{q}$. Hence, all sets in $\mathcal{X}$ can be compared and the one which is most relevant to the question can be identified.

While cosine similarity measures which document set is more related to the question, it does not provide any information on how the document set is focused on the topic(s) of interest. It is possible that  the document set with higher similarity with the question contains much irrelevant information. To explore that property we propose the idea of Document Disparity as an indicator of Coherence:

\begin{definition} (\textbf{Coherence} ${\mathcal C}$).
	Let $D_i$ be a collection of documents written as distributions over topics. We  then define the \textit{Document Disparity} to be the JR divergence of this set of documents and the reciprocal of the document disparity will be the corresponding \textit{ Coherence}:
	\begin{equation}
	\mathcal {C}(D_i)=\frac{1}{DD(D_i)} =\frac{1}{ JR_\alpha^{\boldsymbol \omega}(g({\bf d}_{1}^{(i)}),\dots,g({\bf d}_{N_i}^{(i)})))}.
	\label{eq:coherence}
	\end{equation}

	\label{coherence}
\end{definition}
\vspace*{-0.5\baselineskip}

This function, $\mathcal{C}:\mathcal{X}\to R^+$ is independent of the question $\bf{q}$,  and measures how ``different" the documents are within their set. A lower disparity measure within a set of documents is equivalent to saying that they are focused on the same category of topics, namely, more coherent. 

In our experiments below we will prove that the metrics defined above are contractions and have a low sensitivity number as defined in section \ref{sens}. In general, any relevant DRL should enjoy these properties. 

%To test the sensitivity of the DRL metrics, we calculate DRL measures with different LDA model 200 times and analyze the mean value with variance plot.
%
%In our case, we have the following function: 
%
%\[LDA(x|model):\mathcal{W}\to \mathcal{S}\]
%
%\noindent where $ LDA(x|model)$ is a $K$ dimensional vector which corresponds to the topic proportion representation of the document $x\inn \mathcal{W}$, for the given LDA model. 

%
%In natural language scenario, it is not easy to quantitatively define the noisiness of a piece of text. However, certain preprocessing steps such as removing stopwords, filtering out non-character symbols, spelling correction, and word normalization, etc. are commonly used to clean up the text for further analysis. We will use a similar heuristic way to clean up the collection of documents

%% file: Methodology.tex
\section{Methodology}\label{methodology}
\begin{figure}[ht!]
\vspace*{-0.1\baselineskip}
	\begin{center}
	\includegraphics[width=0.4\textwidth]{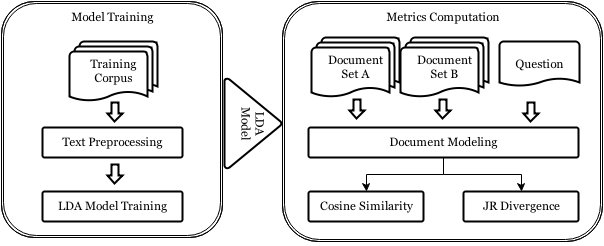}
	\vspace*{-0.5\baselineskip}
	\caption{Methodology flow chart}
	\label{fig:DRL_flow_chart}
\end{center}
\vspace*{-1.5\baselineskip}
\end{figure}

Given a set of document collections $\mathcal X=\{D_1,D_2, ..., D_M\}$, Figure \ref{fig:DRL_flow_chart} shows our framework to compute the aforementioned DRL metrics. Our method is comprised of two parts: the first involves a model training process, while the second evaluates the metrics for relevance and coherence using the previous definitions of relevance and coherence. 

%We expect the most valuable set of data with regard to our question $\bf{q}$ could be selected out with these metrics. 
%The underlying assumption is a valuable piece of data $X$ should have high similarity  and low disparity, indicating high Relevance and high Coherence respectively.  

In  model training, a training corpus ${D}\subset \mathcal{X}$ is preprocessed to build an LDA model.  Let $V$ be the number of words in the vocabulary. According to the BOW assumption, each document $d\in D$ is a $V$-dimensional vector $ d\in \mathbb{R}^V$, a point in the so-called ``word space''. The vector is sparse because only a few words in the vocabulary will appear in the document. The ``LDA Model Training'' module takes as inputs the vector representation of the training corpus $D$ and the number of topics $K$, and generates an LDA model, $g_D:\mathcal{X}\to \mathcal{S}$, that could map a single document ${\mathbf d}$ from a ``word space'' of $V$ dimension to a ``semantic space'' of $K$ dimension. Note here that the training corpus $D$ does not need to be the set of document collections $\mathcal X$. Other than the LDA model, other representation learning approaches such as doc2vec \cite{le2014distributed} and  topic modeling methods \cite{blei2012probabilistic} could be applied to approximate the mapping to the semantic space. Since we are more interested in analyzing the results of DRL, we will leave the analysis of the best document vector representation for the future. 

The second part of the process is the computation of the defined metrics as follows. Given a set of document collections $\mathcal X$ and a specific question $\mathbf{q}$, we project each document $\mathbf{d}\in D_i$, $D_i\in \mathcal X$, $i=1,2,\dots, M$ and the question  $\mathbf{q}$ to the semantic space $\mathcal S$ using the learned function $g_D$. Both cosine similarity $Sim(D_i,\mathbf{q})$ and coherence $C(D_i)$ are computed in the semantic space for all $D_i\in \mathcal X$ with Eq.\ref{eq:relevance} and Eq.\ref{eq:coherence} respectively. %Noisiness dimension is handled implicitly through the document modeling procedure and affects the calculation of relevance and coherence metrics. 

%In text scenario, the noise could be considered as wrongly spelled words, non-character symbols, and even insignificant words based on specific problem settings. 

%% file: Experiments.tex
\section{Experiment and Results}
\label{experiment}

\subsection{Dataset Collection}
To validate our proposed DRL measures, we carried out the following experiments using Twitter data that we collected. Specifically, the dataset consists  of a set of 463,790 tweets collected from Twitter's Streaming APIs\footnote{https://stream.twitter.com/1.1/statuses/filter.json} during the FIBA Basketball Word Cup 2014. The collected tweets cover a time span from August 30, 2014, to September 18, 2014.

To circumvent structural difficulties associated with Twitter Data when converting text into vectors, we performed the following pre-processing steps: stopwords such  as ``the'',``of'' and ``and'' were deleted; HTTP links were deleted, leaving only digits and Latin characters; words of document frequency smaller than 5, and all 1 word-tweets were deleted; text was lowercased. 
%\begin{enumerate}
%\setlength\itemsep{-0.3em}
%	\item All stop-words were deleted, including common words such  as ``the'',``of'',``and''.
%	\item All HTTP links were deleted, leaving only digits and Latin characters.
%	\item A filtering of the texts was carried out by deleting all words of document frequency smaller than 5, and all 1 word-tweets.
%	\item A normalization of the text was carried out by converting all text to lower case.
%\end{enumerate}

The cumulative number of  463,204 tweets  yielded a vocabulary size of 12,207. The corpus of tweets was divided into daily sets resulting in a series  $\mathcal X=\{D_1,D_2, ..., D_{20} \}$. The primary objective of the defined experiment was the discovery of the most relevant set to the following question: \textit{Will the USA Basketball team win the world cup in Spain?}. We partitioned the tweets to match the FIBA basketball world cup daily schedule. The assumption underlying our experiment is there being an uptick in the messages of the day's game so that the tweet messages in the days approaching the final should strengthen the relevance of the set to the initial question of interest, namely the ultimate ``win of the cup by the USA''. All tweets about unimportant (or of least relevance) games on a given day are effectively noisier. Conversely, all Team USA games around key dates  (preliminary rounds, knockout games etc.) lead to a large number of daily tweets, which in turn reflect the fans' opinion-prognosis about the question at hand.

\subsection{Experiment Design}
Each tweet was regarded as a document; note that for simplicity, no tweet pooling schemes \cite{Meh:13, Hon:10} (aggregating tweets into one document) were used  (albeit  another  viable alternative to consider). All the tweets were used as the training corpus to train the LDA model. After some experimentation, we converged on 50 as the number of topics of choice, as it gave a set of clear patterns; it is worth noting that this did not greatly impact the final results, but presented some computational advantage. This number is also a key parameter for the LDA algorithm, which for the most part represents the heaviest computational load in the experiment. %A ``sufficient coherence'' principle, currently achieved by trial-and-error is discussed in  \cite{Newm:10,Cha:09}. 
For the actual implementation we used the Gensim Python Library %\footnote{http://radimrehurek.com/gensim/} 
\cite{Reh:10} and run a Standard LDA on the corpus.

For the metrics computation, we first used LDA to project the collection of daily tweets as well as the question to the semantic space. We then computed both the {\em cosine similarity}  between the tweets and the question (as distributions over topics ) as well as the {\em Coherence} of each set. %The application of the LDA model reduced the dimension of the feature space by 99.59\% in this case, thus implying that there is little gain  in computing the similarity measure using the vector space model, as each sample vector is too sparse to reveal their relationship. 
In order to account for the non-deterministic nature of LDA, we performed $n=200$ random computations and have retained the average of the cosine similarity and JR divergence.

\subsection{Experiment Results}
\begin{figure}[ht!]
\vspace*{-0.8\baselineskip}
	\centering
	\includegraphics[width=0.9\columnwidth]{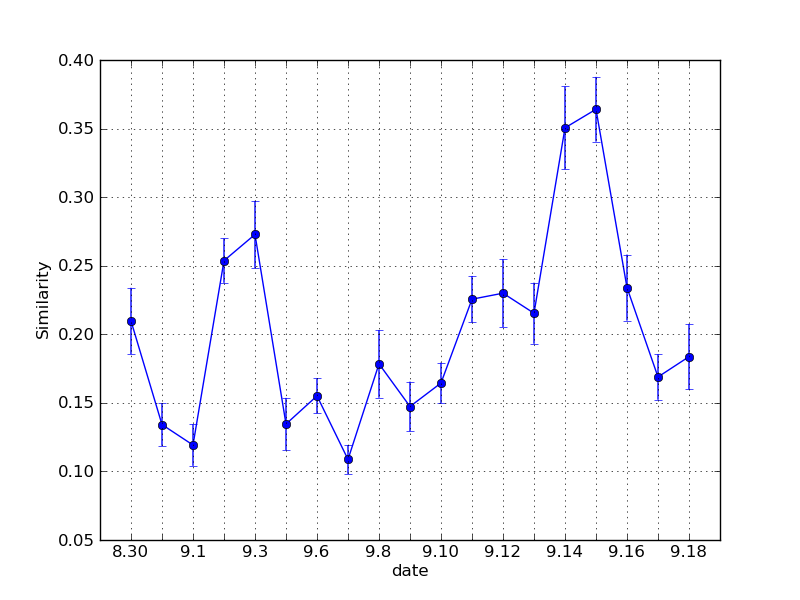}
	\vspace*{-\baselineskip}
	\caption{Daily Cosine Similarity}
	\label{fig:csims}
	\vspace*{-\baselineskip}
\end{figure}

\begin{figure}[ht!]
	\centering
	\vspace*{-\baselineskip}
	\includegraphics[width=0.9\columnwidth]{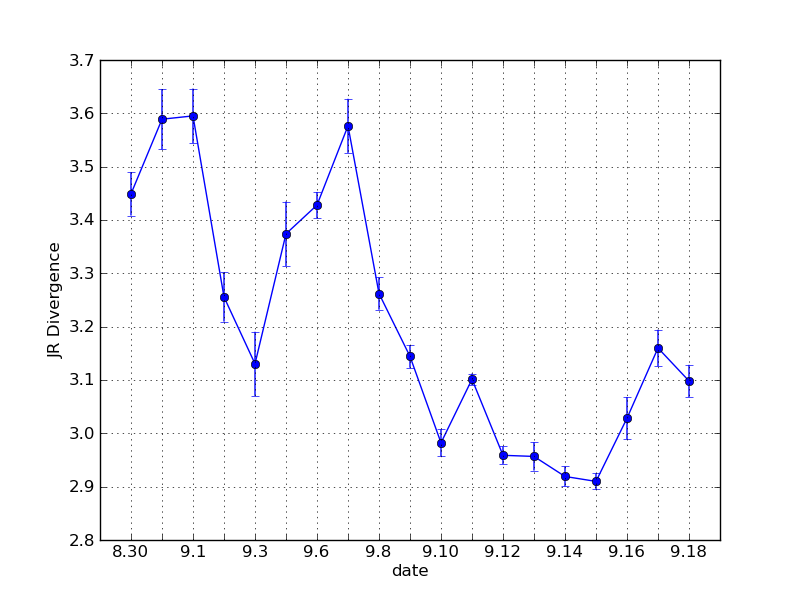}
	\vspace*{-\baselineskip}
	\caption{Daily Document Disparity}
	\label{fig:jrdivs}
	\vspace*{-\baselineskip}
\end{figure}

Our experiment results with the FIBA basketball world cup tweet data, are shown in  Figures \ref{fig:csims} and \ref{fig:jrdivs}. We first note the emergence of a time-lag of at most one day in our results, which is likely due to the difference of  game  occurrence time and the tweet reaction time.

\begin{figure}[ht!]
	\vspace*{-1.6\baselineskip}
	\centering
	\includegraphics[width=0.8\columnwidth]{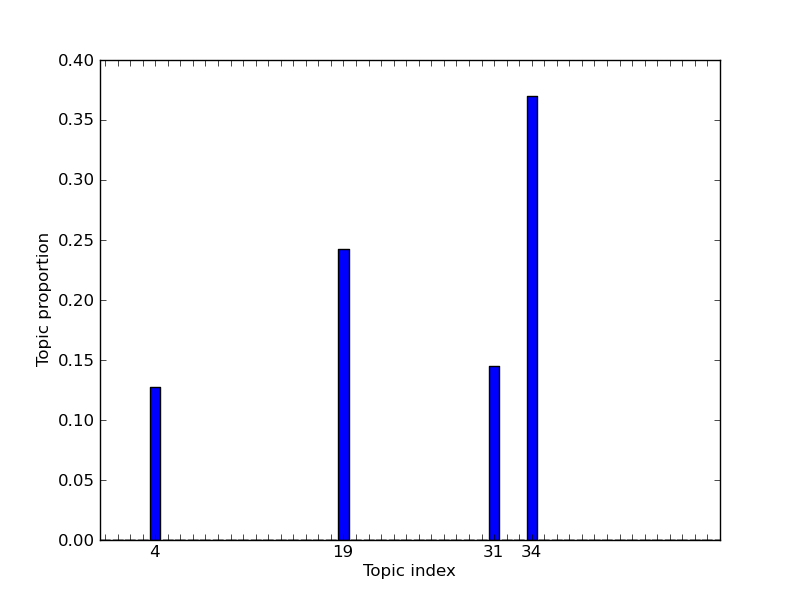}
	\vspace*{-\baselineskip}
	\caption{Topic proportion for the query}
	\label{fig:topic_proportion}
	\vspace*{-\baselineskip}
\end{figure}

The question,  ``Will the USA Basketball team win the world cup in Spain?", was represented as a distribution over topics as shown in figure \ref{fig:topic_proportion}. 
The corresponding topics (top 5 words) were:
\begin{description}
\small
\vspace*{-0.5\baselineskip}
\setlength\itemsep{-0.3em}
	\item[Topic 4]:  0.272*win + 0.119*rose + 0.102*derrick + 0.085*point + 0.083*performance;
	\item[Topic 19]: 0.343*world + 0.318*cup + 0.301*basketball + 0.008*group + 0.006*liked;
	\item[Topic 31]: 0.203*spain + 0.156*world + 0.066*worl + 0.060*celebration + 0.056*cup;
	\item[Topic 34]: 0.287*usa + 0.228*team + 0.211*world + 0.192*cup + 0.011*home.
\end{description}

Figure \ref{fig:csims} shows the average cosine similarity measure for the daily sets of tweets with the question  $\textbf{q}$,  as well the variance over $200$ iterations. We quickly note  that the dates $9.2-9.3$, $9.11-9.12$ and $9.14-9.15$ show relatively high cosine similarity, with that of $9.14-9.15$ being the highest possible. According to the FIBA basketball calendar, USA qualified for the second round on $9.3$, while mathematically on $9.2$, thus explaining the high relevance of the tweets of that day to USA's team. We also note that  on  the $11^{th}$ of September, USA beat Lithuania during the semi-finals game and qualified for the final. The day of the final, $9.14$, is clearly the day when the relation of the tweets is as close as possible to our question, thus revealing in the process the highest output that day.

While one would expect a smaller spike in our graph on $9.9-9.10$, when USA beat Slovenia in the quarterfinals, another concurrent event (game of Lithuania versus Turkey) came in to spread the expected  focus of the day on Team USA, caused by  additional tweets about this other game.
This amounts to declaring  the dataset at that point  ``less ready" to answer the question of interest since more than one topic is discussed that day.
This is more easily seen in the Document Disparity figure \ref{fig:jrdivs} where the $DD$ for $9.9$ is one of the highest recorded, so the coherence being the reciprocal of that is the lowest.

Looking now at figure \ref{fig:jrdivs}, we can see that the lowest document disparity is found on $9.14-9.15$. Again this is easy to explain since the ``Finals'' is an  event of central importance, and would hence be central to the tweets.   We hasten to also point out that  during the first few days $(8.30-9.2)$,  a very high $DD$ was due to the period of preliminary rounds when noise topics were emerging (at least one for each team participating). This trend carried on  to $9.9-9.10$, at the conclusion of  the quarter-finals, when  the discussed topics are now more concentrated on the remaining teams and their upcoming final games. We believe the small dip in the disparity appearing on $9.2-9.3$ is related to the peak of the cosine similarity for those same days, and most likely on account of the fact that Team USA appeared set to qualify for the finals, and possibly a repeat victory. Various tweets of that day make predictions about the winners of the final and mention the fact that the USA is going to win again.

In light of the above results, we believe that  both  measures present a good foundation for DRL, by inferring  relations of the data set to a question of interest. As a result, a data scientist may opt to bypass other sets of documents and focus on the ones with high cosine similarity and small document disparity. Conversely, one may argue, that for a certain class of questions, one should instead select a high cosine similarity and a high document disparity to reflect the highly exploratory nature of a posed question, rather that fact-driven. 
The proposed method, provides in both case, a step towards an automated, semi-supervised method to measure the readiness of a given data set. It remains that a true quantitative DRL is expected to be some {\em  function} of the two metrics discussed above.

\subsection{Perturbation Test}

Our goal is to show that any little  perturbation of  the query, $\mathbf{q}$, will only minimally impact the LDA model (same projection function) to hence provide a relatively consistent semantic representation, i.e. a small $r$-sensitivity number $s_1(\mathbf{q})$.
Given the clear difficulty of  computing the infimum over all possible perturbations of the query, we provide the sensitivity number for various different cases of perturbation and define  a corresponding  sensitivity quotient:
\vspace*{-0.5\baselineskip}
%\begin{align*}
%s_1(\mathbf{q},\mathbf{q_p}) &=\frac{ \text{relative change in semantic representation}}{\text{relative change in query}}\\
%&=\frac { ||g_D(\mathbf{q} ) - g_D(\mathbf{q_p})|| \div \|g_d(\mathbf{q})\|}{ ||\mathbf{q} - \mathbf{q_p}||\div ||\mathbf{q}||}\\
%&=\frac { ||g_D(\mathbf{q}) - g_D(\mathbf{q_p})||\cdot ||\mathbf{q}|| }{ ||\mathbf{q} - \mathbf{q_p}||\cdot ||g_D(\mathbf{q})|| }.
%\end{align*} 
\begin{align*}
s_1(\mathbf{q},\mathbf{q_p}) &=\frac{ \text{relative change in semantic representation}}{\text{relative change in query}}\\
&=\frac { ||g_D(\mathbf{q}) - g_D(\mathbf{q_p})||\cdot ||\mathbf{q}|| }{ ||\mathbf{q} - \mathbf{q_p}||\cdot ||g_D(\mathbf{q})|| }.
\end{align*} 
%\begin{align*}
%s_1(q,q_p) &=\frac{ \text{relative change in semantic representation}}{\text{relative change in query}}\\
%    &=\frac { ||LDA(q | model) - LDA(q_p | model)|| \div ||LDA(q | model)||}{ ||q - q_p||\div ||q||}\\
%    &=\frac { ||LDA(q | model) - LDA(q_p | model)||\cdot ||q|| }{ ||q - q_p||\cdot ||LDA(q | model)|| }
%\end{align*} 

In section \ref{Prereq} we defined the first DRL measure of a set of documents $D$ with respect to a question $\mathbf{q}$ to be $Sim(D,\mathbf{q})$. %Notice that  
%\[DRL:\mathcal{S}\to \mathbb{R}^+\]\
%More precisely,  we view DRL as a measure on the space of documents (as bags of words), which is related to the LDA model in use,  and the question as follows: \[ DRL(D|\mathbf{q})=Sim(LDA(D;model),LDA(\mathbf{q};model)),\] in other words $DRL=(Sim\circ LDA)(X,\mathbf{q};model)$. 
The goal is to then see how sensitive the DRL measure is to a change in the query. Testing this sensitivity number for a set of documents with high DRL(tweets from Sept.15) and another with low DRL (tweets from Sept.7) yields, much like $s_1(q,q_p)$, the normalized sensitivity quotient as, 
\vspace*{-0.5\baselineskip}
%\begin{align*}
%s_2(X,\mathbf{q},\mathbf{q_p}) &= \frac{\text{ relative change in DRL}}{ \text{relative change in query}}\\
%&= \frac{ |Sim(X,\mathbf{q}) - Sim(X,\mathbf{q_p})| \div |Sim(X,\mathbf{q}))|} { ||\mathbf{q} - \mathbf{q_p}||\div ||\mathbf{q}||}\\
%&= \frac{ |Sim(X,\mathbf{q}) - Sim(X,\mathbf{q_p})| \cdot ||\mathbf{q}||} { ||\mathbf{q} - \mathbf{q_p}||\cdot |Sim(X,\mathbf{q}))| }.
%\end{align*}
\begin{align*}
s_2(X,\mathbf{q},\mathbf{q_p}) &= \frac{\text{ relative change in DRL}}{ \text{relative change in query}}\\
&= \frac{ |Sim(X,\mathbf{q}) - Sim(X,\mathbf{q_p})| \cdot ||\mathbf{q}||} { ||\mathbf{q} - \mathbf{q_p}||\cdot |Sim(X,\mathbf{q}))| }.
\end{align*}

We have constructed  various perturbations of the query based on word repetition, synonyms replacement, and word deletion. Having fixed a dictionary, the bag of words representation of the query is similarly independent of the set of documents $\mathcal{X}$, all the while using the  LDA model. Note though that the semantic representation through the LDA, heavily depends on $\mathcal{X}$.

%The query for this experiment is, \textit{Will the USA Basketball team win the world cup in Spain?}  and is represented by:
%\begin{align*}
%[&(280,1),&(773,1),&(1291,1),&(2717,1),&(5307,1),&(8167,1),&(11527,1)]\\
%[&'basketball', &'team', &'world', &'Spain', &'win',&'cup', &'USA']
%\end{align*}
%where $(280,1)$ means the word ('basketball') with index $280$ in the dictionary occurs once. By word repetition, we simply repeat one word in the query. For example, if we repeat word \textit{USA}, the bag of words representation becomes:
%\begin{align*}
%[&(280,1),&(773,1),&(1291,1),&(2717,1),&(5307,1),&(8167,1),&(11527,2)]\\
%[&'basketball', &'team', &'world', &'Spain', &'win',&'cup', &'USA']
%\end{align*}
%By synonyms replacement, we mean to replace  one word with one of its synonyms. For example, if 'Spain' is replaced by 'Spain2014', the query becomes:
%\begin{align*}
%[&(280,1),&(351,1),&(773,1),&(1291,1),&(5307,1),&(8167,1),&(11527,1)]\\
%[&'basketball', &'Spain2014', &'team', &'world', &'win',&'cup', &'USA']
%\end{align*}
%By word deletion we mean the removal of one or two words. For example if we remove the word ``basketball" our perturbed query is:
%\begin{align*}
%[&(773,1),&(1291,1),&(2717,1),&(5307,1),&(8167,1),&(11527,1)]\\
%[&'team', &'world', &'Spain', &'win',&'cup', &'USA']
%\end{align*}

\begin{table}[ht!]
\small

	\centering
	\caption{Perturbed query examples}
	\vspace*{-0.8\baselineskip}
	\begin{tabular}{l}
		\hline
		Repetition\\
		\hline
		 $q_{a1}$: USA USA basketball team win world cup Spain \\
		$q_{a2}$: USA basketball basketball team win world cup Spain \\
		$q_{a3}$: USA basketball team team win world cup Spain \\
		$q_{a4}$: USA basketball team win win world cup Spain \\
		 $q_{a5}$: USA basketball team win world world cup Spain \\
		 $q_{a6}$: USA basketball team win  world cup cup Spain \\
		 $q_{a7}$: USA basketball team win world cup Spain Spain \\
		\hline
		Replacement\\
		\hline
		$q_{b1}$: USA basketball team win FIBA Spain \\
		$q_{b2}$: USA basketball team win world cup FIBA \\
		$q_{b3}$: USA basketball team win  world cup Spain2014 \\
		\hline
		Deletion\\
		\hline
		 $q_{c1}$: USA team win  world cup Spain \\
		 $q_{c2}$: USA basketball  win world cup Spain \\
		 $q_{c3}$: USA basketball team win world cup \\
		 $q_{c4}$: USA basketball team win Spain \\
		\hline
	\end{tabular}%
	\label{tab:perturbed_query}%
	%\vspace*{-\baselineskip}
\end{table}%

\begin{figure}[ht!]
\vspace*{-0.5\baselineskip}
	\begin{center}
		\begin{subfigure}[b]{0.3\columnwidth}
			\includegraphics[width=\columnwidth]{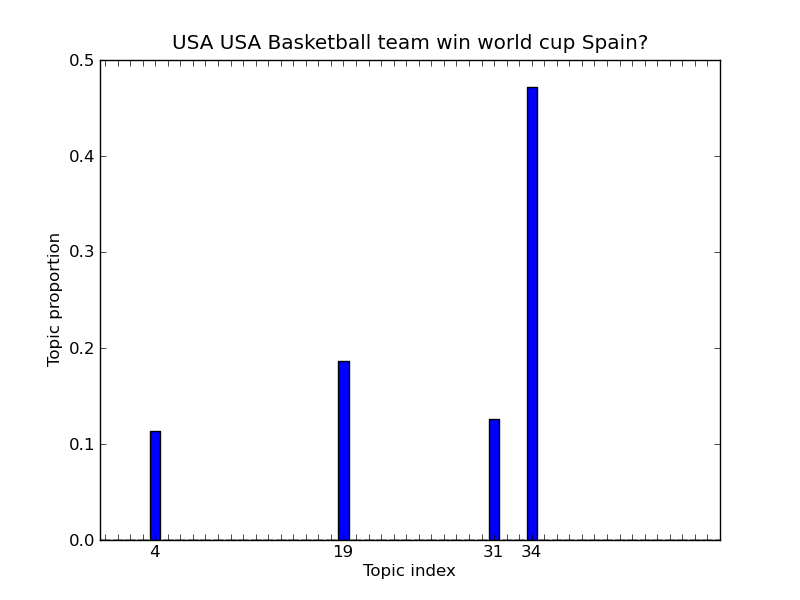}
			\caption{$q_{a1}$}
			\label{fig:qp_rep1}
		\end{subfigure}%
		~ %add desired spacing between images, e. g. ~, \quad, \qquad, \hfill etc.
		\begin{subfigure}[b]{0.3\columnwidth}
			\includegraphics[width=\columnwidth]{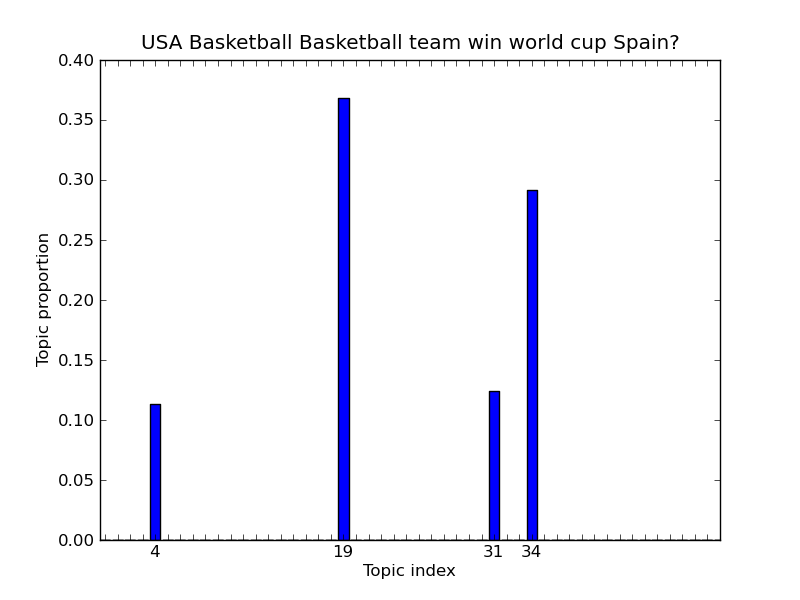}
			\caption{$q_{a2}$}
			\label{fig:qp_rep2}
		\end{subfigure}
		~ %add desired spacing between images, e. g. ~, \quad, \qquad, \hfill etc.
		\begin{subfigure}[b]{0.3\columnwidth}
			\includegraphics[width=\columnwidth]{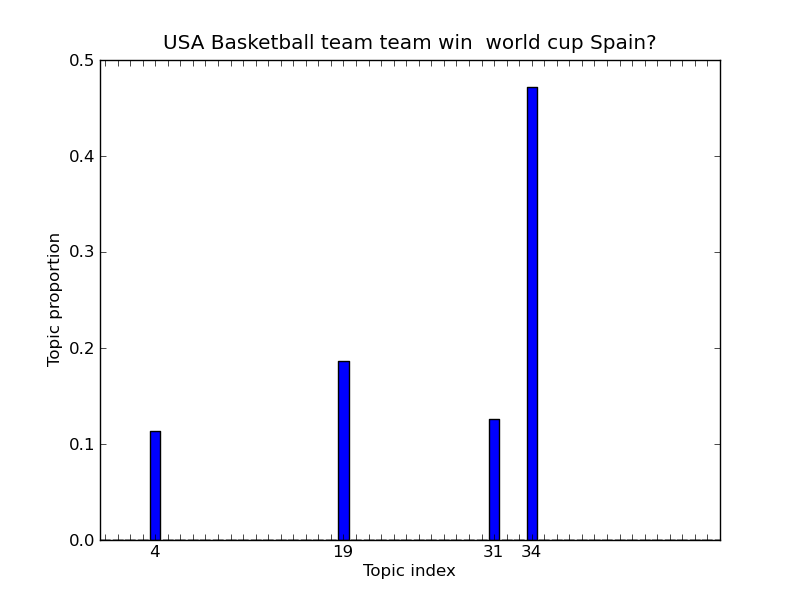}
			\caption{$q_{a3}$}
			\label{fig:qp_rep3}
		\end{subfigure}
		
		\begin{subfigure}[b]{0.3\columnwidth}
			\includegraphics[width=\columnwidth]{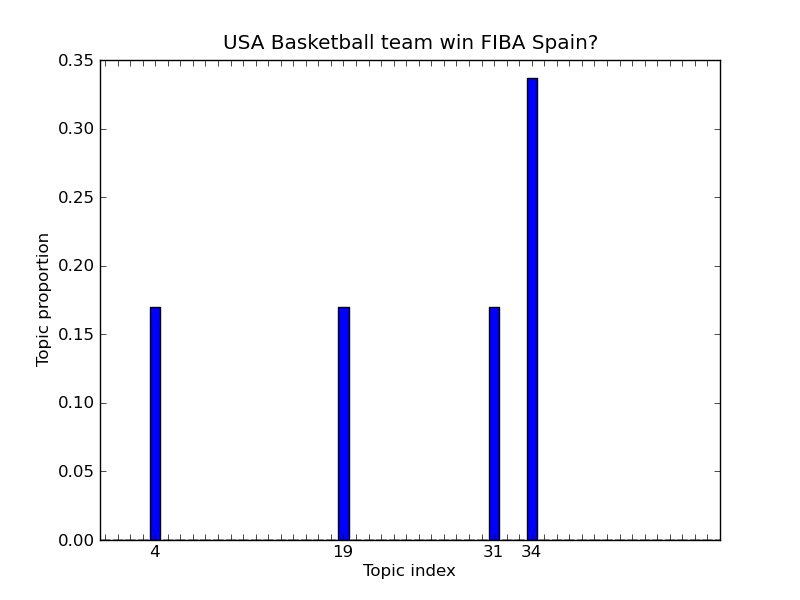}
			\caption{$q_{b1}$}
			\label{fig:qp_rpl1}
		\end{subfigure}%
		~ %add desired spacing between images, e. g. ~, \quad, \qquad, \hfill etc.
		\begin{subfigure}[b]{0.3\columnwidth}
			\includegraphics[width=\columnwidth]{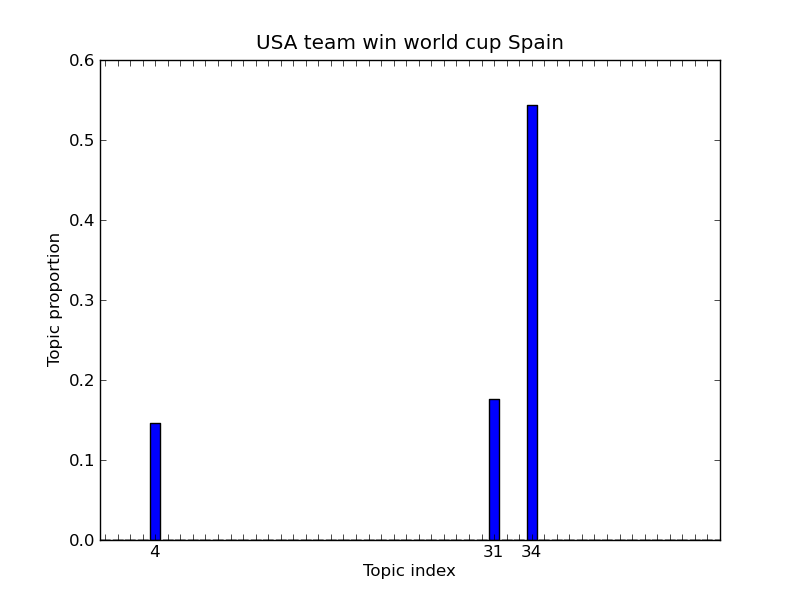}
			\caption{$q_{c1}$}
			\label{fig:qp_del1}
		\end{subfigure}
		~ %add desired spacing between images, e. g. ~, \quad, \qquad, \hfill etc.
		\begin{subfigure}[b]{0.3\columnwidth}
			\includegraphics[width=\columnwidth]{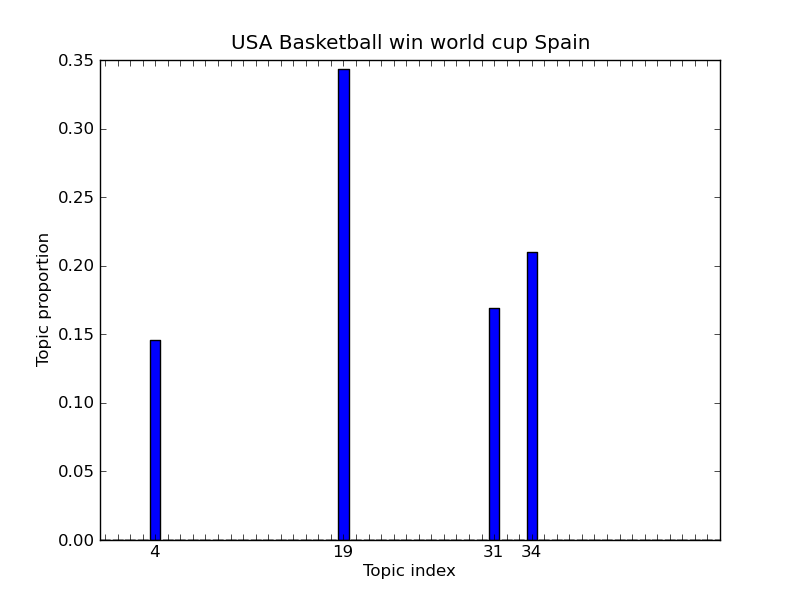}
			\caption{$q_{c2}$}
			\label{fig:qp_del2}
		\end{subfigure}
	\end{center}
	\vspace*{-\baselineskip}
	\caption{Topic Distributions of Perturbed queries}\label{fig:qp_topic_distribution}
	%\vspace*{-\baselineskip}
\end{figure}

\begin{figure}[ht!]
\vspace*{-0.5\baselineskip}
	\centering
	\includegraphics[width=0.9\columnwidth]{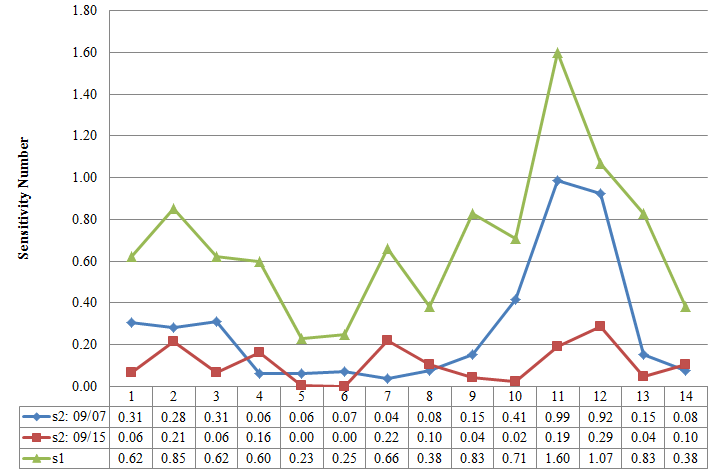}
	
	\caption{Sensitivity numbers}
	\label{fig:sensitivity_numbers}
	\vspace*{-1.5\baselineskip}
\end{figure}

A list of perturbed queries is shown in table \ref{tab:perturbed_query}. The topic distributions of some of these perturbed queries are shown in Figure \ref{fig:qp_topic_distribution}. The sensitivity numbers are shown in Figure \ref{fig:sensitivity_numbers}: The entries $1-7$ correspond to the queries $q_{a1}-q_{a7}$. The entries $8-10$ correspond to the queries $q_{b1}-q_{b3}$ and the entries $11-14$ to the queries $q_{c1}-q_{c4}$. 

In almost all cases, the sensitivity quotient number $s_1$, corresponding to ``LDA" in the figure, is smaller than $1$. Only for the cases $q_{c1}$ and $q_{c2}$ is it bigger than one. This is because topics 19 (about the world cup basketball) and 34 (about the USA basketball team) were ill-proportioned in the perturbed query due to loss of information. One can suggest that this change (deletion of one word) is small in terms of distance in $\mathcal{W}$ but relatively large in the semantic space $\mathcal{S}$. The sensitivity quotient number $s_2$ is also smaller than $1$. 

%\pagebreak

%% file: Relatedwork.tex
\section{Related Work}\label{Relatedwork}

The idea of Data Readiness Levels is closely related to the field of information retrieval. The main goal of traditional information retrieval is to find relevant documents to a natural language query \cite{zhai2008statistical}. The desired output there is for the most relevant documents to be top-ranked in the returned list of documents. In our research, though, we would like to account for other data modalities such as videos and images. The general definition of DRL would evaluate the value of a piece of data to an objective while information retrieval measures only the relevance of a piece of information to a query. It is true that retrieved information based on the query should have a higher DRL than a randomly selected piece of data with respect to the same query. In this case, information retrieval techniques, especially question answering, could be applied to increase the DRL of data in the relevance dimension and probably coherence dimension. In cases where text is the main data medium, DRL is a different problem than information retrieval in that DRL focuses on the ``goodness" of a collection of documents to an objective instead of a single document. 

Technology Readiness Level (TRL) can be thought of as a template for DRL. TRL is a well-established methodology of estimating the technology maturity during the acquisition process to assist decision-making concerning technology funding and transition \cite{dod2011technology}. The U.S. Department of Defense (DoD) has defined TRL based on a scale from 1 to 9 with 9 being the most mature technology. Unlike TRL, there are no standard evaluation rules for data maturity but we propose that it is not possible to define a general DRL without a specific objective. For example, a collection of documents about basketball games has less value to answer a question related to volleyball than to basketball no matter how refined these documents are. 

%% file: Conclusion.tex
\section{Conclusions}\label{conclusion}
In summary, cosine similarity measures the average relation of a document set and a given question, while document disparity reflects the variance or diversity of the information contained in a document set. Given the two measures, we would favor a data set with larger $Sim$ measure and lower $DD$ measure. Such a data set would have a high data readiness level. 

By a combined theoretical-experimental approach, we have laid out some groundwork  for a viable definition of a computable  DRL measure, usable for any analytic process, especially for large data sets. While heavily intuited and illustrated using document-based data, DRL should be a generic and flexible measure compatible with any data modality. A great deal of work clearly remains to be undertaken, particularly in accurately establishing a quantitative scale for it, as well as applying to other modalities such as images, audio signals etc.

%Our work, thus far, has not delved much into the semantics information realm where much remains to be done. Another aspect which we would contemplate exploring in the futures is the dynamic nature of the  DRL measure when one can use  ideas from information retrieval \cite{Cro:09} and natural language processing \cite{Bat:95}. These will be especially appropriate when the text documents are more structured and richer in content.  

We believe that by exploring the idea of the semantic space, a concept which is not very well understood, the linking between various modalities and a specific question will be achieved, opening the way for comparison and computations on readiness. Here we are not using the standard ideas behind the semantic space as the re-formulation of documents through a topic discovery model. What we really mean is the creation of an abstract space, containing cognitive information and associations between ideas, similar to human intellect and understanding. In this semantic space, the concepts contained in documents, images, and other signals, will have a common reference system enabling us to link and compare them. Distances in this semantic space will formalize the idea of ``notionally close" that humans naturally possess.

%% file: Acknowledgement.tex
\section*{Acknowledgment}
We would like to thank all the members of the Data Readiness Level Team especially Dr. Harish Chintakunta, for meaningful discussions on the subject, and the Laboratory for Analytics Sciences for its generous financial support.